\documentclass[a4paper,useAMS,usenatbib]{mn2e}

\usepackage{graphicx}
\usepackage[fleqn]{amsmath}
\usepackage{amssymb}
\usepackage{color}
\usepackage{url}

\title[The formation of Kepler-36]{The formation of systems with closely spaced low-mass planets and the application to Kepler-36}

\author[Paardekooper et al.]{Sijme-Jan Paardekooper$^1$\thanks{E-mail: 
\texttt{S.Paardekooper@damtp.cam.ac.uk}}, Hanno Rein$^{2,3}$, Willy Kley$^4$\\  
$^1$DAMTP, University of Cambridge, Wilberforce Road, Cambridge CB3 0WA,
United Kingdom\\
$^2$Institute for Advanced Study, 1 Einstein Drive, Princeton, NJ 08540\\
$^3$University of Toronto, Scarborough, 1265 Military Trail, Toronto, ON M1C 1A4, Canada\\
$^4$Institut f\"ur Astronomie \& Astrophysik, Universit\"at T\"ubingen, Auf der Morgenstelle 10, 72076 T\"ubingen, Germany}

\begin{document}

\date{Draft version \today}

\pagerange{\pageref{firstpage}--\pageref{lastpage}} \pubyear{2013}

\maketitle

\label{firstpage}

%------------------
% Abstract text
%------------------

\begin{abstract}
The Kepler-36 system consists of two planets that are spaced unusually close together, near the $7$:$6$ mean motion resonance. 
While it is known that mean motion resonances can easily form by convergent migration, Kepler-36 is an extreme case due to the close spacing and the relatively high planet masses of 4 and 8 times that of the Earth. 
In this paper, we investigate whether such a system can be obtained by interactions with the protoplanetary disc. 
These discs are thought to be turbulent and exhibit density fluctuations which might originate from the magneto-rotational instability. 
We adopt a realistic description for stochastic forces due to these density fluctuations and perform both long term hydrodynamical and $N$-body simulations. 
Our results show that planets in the Kepler-36 mass range can be naturally assembled into a closely spaced planetary system for a wide range of migration parameters in a turbulent disc similar to the minimum mass solar nebula. 
The final orbits of our formation scenarios tend to be Lagrange stable, even though large parts of the parameter space are chaotic and unstable.
\end{abstract}

%---------------
% Keywords
%---------------
 

%-----------------------
% Table of Contents
%-----------------------
%
%\tableofcontents
%

%-----------------------
% Introduction text
%-----------------------

\section{Introduction}
\label{secInt}

Over the past two decades, more than $800$ extrasolar planets have been discovered\footnote{See e.g. \url{http://openexoplanetcatalogue.com}}, and more than $100$ multi-planetary systems. The most striking characteristic of this collection of planets is the enormous variety, for example in semi-major axis, ranging from Hot Jupiters like 51 Peg \citep{mayor95}, with periods on the order of days, to giant planets on orbits of more than 500 years like in HR\,8799 \citep{marois08}. Planets have been detected in extreme environments: in close binary systems, e.g. $\gamma$-Cep \citep{neuhauser07}, in circumbinary orbits, e.g. Kepler-16 \citep{doyle11}, and around pulsars \citep{wolszczan92}. A theory of planet formation somehow has to be able to explain these extreme planetary systems. Or, viewed from a different angle, extreme planetary systems can be useful testbeds for various theories of planet formation. 

A recent addition to the class of extreme planets came in the form of Kepler-36 \citep{carter12}: a pair of planets of dissimilar densities spaced extremely close together. 
The inner planet is a Super Earth, while the outer planet is a small Neptune-type planet. 
Their semi-major axis differ only by approximately 1\% of an astronomical unit (AU), so that the outer planet on closest approach appears twice as large as the full moon as viewed from the inner planet. 
In fact, the planets are close to, but just outside, a $7$:$6$ mean motion resonance \citep[MMR,][]{carter12}. 
For convenience, we summarise the main properties of the Kepler-36 system in Table \ref{tabKepler}.

The existence of a closely-packed planetary system such as Kepler-36 raises some important questions about planet formation. 
How is it possible that systems like Kepler-36 form in such a compact configuration? While the orbital configuration of Kepler-36 is not special in terms of period ratio (it is not exactly in the $7$:$6$ MMR), it appears to be special in terms of stability. 
As found by \cite{deck12}, the system is very close to unstable regions in parameter space. 
For most of the orbital solutions the Lyapunov timescale is of the order of a few hundred years, compared to millions of years in our own Solar System \citep{Hayes2007}.
This means the system is chaotic, in the sense that the memory of initial conditions is lost on short timescales. 
Nevertheless the system is stable over much longer timescales \citep{deck12}.

How did the Kepler-36 planets arrive exactly in such an island of stability? Their different mean densities suggest instead that they have formed in different locations, which allowed only the outer planet to capture a significant amount of gas from the disc out of which the planets formed.
Their close proximity to the central star suggests that both have migrated inward a significant distance. 
Combining both these suggestions, a possible formation scenario involves convergent migration, a mechanism that is known to lead to resonant pairs of planets, sparked by the discovery of Gliese 876 \citep{marcy01, snellgrove01, lee02}. Convergent migration into the $7$:$6$ MMR could lead to a stable planetary system, and, as suggested by \cite{deck12}, subsequent tidal evolution, either by interaction with the star \citep{papaloizou11, lithwick12, batygin13}, or with the remnant disc \citep{baruteau13}, could drive the planets slightly out of the $7$:$6$ resonance. Note, however, that these mechanisms work in the wrong direction for Kepler-36 as this system is observed to be slightly inward of resonance. Subsequent evolution could also be driven by evaporation of the inner planet's atmosphere \citep{OwenWu2013}.

Formation mechanisms involving convergent migration have difficulties when the inner planet is massive enough to significantly affect the surface density of the disc. 
In that case an outer low-mass planet stops migrating before even reaching the $2$:$1$ MMR due to interaction with the density waves launched by the inner planet \citep{podlewska12}. 
Thus, if both planets are massive, forming resonances as closely spaced as the $4$:$3$ MMR poses problems \citep{rein12}. 
However, since both planets in the Kepler-36 system are in the Super-Earth regime and therefore considered low-mass in terms of their interaction with the disc, and we can be confident that they do not significantly perturb the surface density profile.

Even so, if the planets formed far apart, how did they push through all other resonances on their way to the $7$:$6$? 
This is the question we address in this paper. 
While it is known that for sufficiently high disc masses, low-mass planets can be pushed into compact configurations \citep{papaloizou05}, the need for a very massive disc brings its own problems: the planets would need to have formed very early on, possibly during the self-gravitating phase of the disc, and they subsequently have to survive the fast migration (with a migration timescale of less than a thousand years) associated with massive discs. 

In this paper, we explore the effects of a turbulent disc on the formation of MMRs for pairs of low-mass planets such as the Kepler~36 system. 
If the stochastic kicks experienced by the planets due to turbulent density fluctuations in the disc can break some of the early, widely spaced resonances, compact configurations may be formed in less massive, more realistic discs. 
These stochastic kicks can be caused by different physical mechanisms. 
The most promising one is the Magneto-Rotational Instability (MRI, see Sect.~\ref{secMMRmig}) which is thought to be active in at least some parts of the protoplanetary disc \citep{nelson04}.

We find that such closely spaced configurations are indeed a very natural outcome of planet disc interactions in a standard turbulent protoplanetary disc.
The systems that form in this scenario closely resemble Kepler-36.
Some of the orbits are chaotic on short timescales but almost all are stable over timescales comparable with the age of the system.
These results are remarkably unsensitive to the speed of convergent migration and to the strength of stochastic forces.

The paper is organised as follows. 
We briefly review the physics of planet migration and mean motion resonances in Sect.~\ref{secMMR}. 
We then describe the results of hydrodynamical simulations in Sect.~\ref{secHydro}, and present the results of $N$-body integrations in Sect.~\ref{secNbody}. 
We discuss the stability of the system in Sect.~\ref{secStability}.
We finally summarise, discuss and conclude in Sect.~\ref{secCon}.

%----------------------
% Kepler-36 table
%----------------------

\begin{table}
  \begin{tabular}{lll}
  \hline
    & Planet b & Planet c \\ \hline
Mass ($10^{-5} M_*$) & $1.33_{-0.081}^{+0.099}$ & $2.42_{-0.14}^{+0.18}$ \\
Semi-Major axis (AU) & $0.1153_{-0.0015}^{+0.0015}$ & $0.1283_{-0.0016}^{+0.0016}$ \\
Eccentricity & $<0.04$ & $<0.04$ \\
Period (d) & $13.83989_{-0.0060}^{+0.0082}$ & $16.23855_{-0.0054}^{+0.0038}$ \\
Mean density (g $\mathrm{cm^{-3}}$) & $7.46_{-0.59}^{+0.74}$ & $0.89_{-0.05}^{+0.07}$ \\
\hline
  \end{tabular}
  \caption{Properties of the Kepler-36 planetary system, where the stellar
 mass is $M_* = 1.071~ \mathrm{M_\odot}$. Data adapted from \citet{carter12}.}
  \label{tabKepler}
\end{table}

%-------------------------------------------
% Mean Motion Resonances text
%-------------------------------------------

\section{Planet migration and Mean Motion Resonances}
\label{secMMR}

In this section, we give an overview of the relevant physics of planet migration and the formation of mean motion resonances. 

\subsection{Type I planetary migration}
\label{secMMRmig}

Ever since \cite{goldreich80} it has been known that embedded planets can exchange angular momentum and energy with gaseous discs, leading to changes in their orbit. 
Several types of disc-induced planet migration can be distinguished, according to the nature of the interaction with the disc \citep{ward97}. 

Low-mass planets interact with the disc in a linear fashion, exciting density waves but leaving the overall structure of the disc intact. 
The torque $\Gamma$ on the planet resulting from these density waves can be calculated semi-analytically for an isothermal disc \citep{tanaka02} as
\begin{equation}
\Gamma=-C\frac{q^2}{h^2}\Sigma_p r_p^4\Omega_p^2 \,.
\label{eqTorque}
\end{equation}
Here $q$ denotes the mass of the planet in terms of the mass of the central star, $h$ is the aspect ratio of the disc, $\Sigma$ is the surface density, $r$ is the orbital radius and $\Omega$ is the angular velocity. 
Subscripts $p$ denote that quantities have to be evaluated at the location of the planet. The constant $C$ is of order unity and depends on the local surface density slope, with in general $C>0$ leading to a negative torque on the planet and therefore inward migration. 
Allowing for non-isothermal effects leads to a different $C$ that now depends very strongly on the local temperature gradient, with the possibility of $C<0$ and outward migration \citep{paard06,2008A&A...487L...9K,paard10}.
This migration regime is called Type I. 

For high-mass planets, interaction with the disc becomes non-linear, leading to shocks close to the planet and the opening of an annular gap around the orbit of the planet \citep{lin86}. 
This is known as Type II migration.
The critical parameter determining whether wave-like interactions with the disc are linear is $q/h^3$ \citep{korycansky96}. 
The most massive planet of Kepler-36 has $q/h^3 \approx  0.2$ for the canonical value of $h=0.05$, which means that wave-like interactions with the disc are expected to be linear. 
The Kepler-36 planets will therefore be subject to Type I migration when embedded in a gaseous disc. 

Since the Type I torque is proportional to the mass squared ($q^2$, see Eq.~\ref{eqTorque}), the migration speed is proportional to the mass of the planet. 
Multiple planets embedded in a disc, if they are of different mass and migration is directed inward, will therefore either migrate away from each other (if the inner planet is the most massive) of migrate towards each other (if the outer planet is the most massive). 
The latter case of convergent migration is interesting for the formation of resonances. 

Type I planet migration is usually studied in laminar discs, where a Navier-Stokes viscosity may be included to obtain an accretion flow in the disc. 
In reality, accretion will be driven by turbulence in the disc, most likely due to the MRI \citep[see][]{balbus91}. 
The effects of turbulence on Type I migration can be profound. Turbulent density fluctuations introduce a stochastic component in the torque from the disc felt by the planet \citep{nelson04}. 
In absence of an average migration torque, planets would undergo a random walk through the disc. 
If a smooth inward migration torque is present, planets would still migrate inward on average, but if the stochastic component of the torque can dominate for long enough, the survival probability of planets can be increased \citep{nelson04, rein09, adams09}. 

It is important to note that there are other possible mechanisms that could add a stochastic component to the migration forces even in the absence of the MRI.
For example, planetesimals and small protoplanets can have a similar effect. 
This makes our discussion slightly more general, although the details might vary between different driving mechanisms.

\subsection{Mean motion resonances}
When two or more bodies orbit the same central object, mean motion resonances (MMR) can occur, where
the orbital period ratio of two of the bodies is close to an integer ratio.
Formally, a system is said to be in a $p$:$q$ resonance if at least one of the resonant angles, $\phi_{1,2}$, is librating,
i.e. has a dynamical range smaller than $2\pi$. The resonant angles are defined as 
\begin{equation}
\phi_{1,2} = p \lambda_1 - q \lambda_2 - (p-q) \, \varpi_{1,2},
\label{eqResAngles}
\end{equation}
where $\lambda_{1,2}$ is the longitude of the inner and outer planet, respectively, and $\varpi_{1,2}$ are their pericenters. 
The difference in pericenters $\Delta \varpi \equiv \varpi_1-\varpi_2$ can be expressed as a linear combination of $\phi_1$ and $\phi_2$, and is often
used to characterize resonant behaviour.

Several celestial bodies in our own Solar System are in resonance.
For example the Jovian satellites Io, Europa and Ganymede are engaged in a so called $1$:$2$:$4$ Laplace resonance,
while Neptune and Pluto (and all Plutinos) are located in a $3$:$2$ MMR.
However, out of the 8 major planets in the Solar System not a single pair is presently in a MMR.
According to the so called Nice model of the early Solar System, this might have been different in the past \citep{Tsiganis2005}.

Extrasolar planetary systems, on the other hand, show often evidence for multiple planets in a mean motion resonance.
The best studied system in Gliese~876, a system of massive planets in a $2$:$1$ MMR 
\citep[e.g.][]{marcy01,LeePeale01,lee02,snellgrove01,NelsonPapaloizou2002,beamic2003,veras2007,2008A&A...483..325C,rein09}.
All these studies suggest that a dissipative process that changes the energy (semi-major axis) of the orbits is required to bring the systems to a resonant configuration. The prime mechanism for this type of sculpting of planetary systems is planetary migration 
\citep{2012ARA&A..50..211K}.
Convergent migration has also been shown to be plausible for more closely spaced systems such as the $3$:$2$ MMR \citep[][]{ReinPapaloizouKley2010}, 
while the formation of even more closely spaced $4$:$3$ resonances can pose serious problems \citep{rein12}.

While the protoplanetary disc, and therefore migration forces, are present, the planets move within the disc in a self-similar fashion \citep[see e.g.][]{LeePeale01}.
The planetary systems formed by such convergent migration process typically have both resonant angles $\phi_{1,2}$ in 
small amplitude libration, a state called apsidal corotation.
These planetary systems are in a stable configuration if the planetary masses are not too large.
Rarely, resonant systems formed by migration show signs of an instability \citep{PierensNelson2008}. 
Whether the system becomes unstable or not depends mainly on the mass ratio of the planets \citep{ReinPapaloizouKley2010}.
We did not observe such an instability for Kepler-36 in our migration scenario.

Another possibility to lose a resonance after being formed by migration is the presence of a stochastic force \citep{rein09}. 
In this case the amplitudes of both resonance angles $\phi_{1,2}$ undergo a random walk and grow proportionally to the square root of time. 
The diffusion coefficient of this random walk can be estimated analytically as shown by \cite{rein09} or calibrated to numerical simulations \citep{nelson04,oishi2007}.

Here, we try to form a system in an extremely close, almost orbit crossing, $7$:$6$ resonance. 
Although the system is not formally in resonance, the close proximity suggests that it might have been in resonance at some point in its history or that the resonance played at least an important role in its formation.
These results will give us valuable information about the possible formation paths of this system and the properties of protoplanetary discs during planet formation.

%---------------------------------------------
% Hydrodynamical simulations text
%---------------------------------------------

\section{Hydrodynamical simulations}
\label{secHydro}

We start by describing results from hydrodynamical simulations, in which we evolve the gaseous disc dynamically together with the planetary system. 
 
\subsection{Numerical method}
\label{secNumHydro}

We use the publicly available code {\sc fargo}\footnote{\url{http://fargo.in2p3.fr}} \citep[Fast Advection in Rotating Gaseous Objects,][]{masset00}, which solves the locally isothermal equations of hydrodynamics on a cylindrical grid in two dimensions $(r,\varphi)$. All quantities (surface density $\Sigma$, velocity ${\bf v}$ and pressure $p$) are treated as vertically integrated. As computational domain we choose $0.2<r/r_0<1.5$ with the full $2\pi$ in azimuth. Here, $r_0$ is a reference radius, which we take to be 1 AU. A natural unit of time is then $2\pi/\Omega(r_0)\equiv 2\pi/\Omega_0$, the orbital period at $r_0$. The choice of $r_0=1$ AU is purely for notational convenience, since time will then be in units of years. The solutions do not depend on the particular choice of $r_0$, and the results can easily be adapted to different systems by appropriately scaling the disc mass (see Sect. \ref{secResHydroLam}) and unit of time. However, it is difficult to choose a single value of $r_0$ that describes for example the Kepler-36 system, where the planets are located at $r\sim 0.1$ AU, because the planets are migrating during the simulation. The computational domain is covered with a grid that has $256$ cells in the radial and $768$ cells in the azimuthal direction. This means one pressure scale height is resolved by approximately 10 computational cells, which is enough to resolve the gas flow responsible for planet migration. 

A surface density profile $\Sigma=\Sigma_0 (r/r_0)^{-1/2}$ was adopted, together with a constant kinematic viscosity $\nu=10^{-5} r_0^2\Omega_0$. 
Note that the choice of surface density slope implies no viscous evolution of the surface density profile, which is convenient when following the evolution of the system for time spans comparable to the viscous time scale at the inner edge of the computational domain. 
In this work, we use viscosity primarily for keeping the solution well-behaved over long time scales. 
If we take $r_0$ to be 1 AU, the Minimum Mass Solar Nebula \citep[MMSN,][]{hayashi81} surface density at $r=r_0$ equals $\Sigma_{0,\mathrm{MMSN}}=2\cdot 10^{-4} M_* /r_0^2$.

The disc has a constant aspect ratio $h=0.05$. 
As mentioned in Sect.~\ref{secMMRmig}, this means that the largest planet with mass ratio $q=M_p/M_*=2.4\cdot 10^{-5}$ has $q/h^3=0.2$, which means disc-planet interactions are well inside the linear regime as far as wave launched by the planets are concerned. 
Corotation torques can be nonlinear for any planet mass \citep{drag}, but by taking a substantial viscosity together with a locally isothermal equation of state we aim at making disc-planet interactions as simple as possible for this study. 
By taking a locally isothermal equation of state, we suppress any complications from corotation torques due to radial gradients in entropy \citep{paard08, baruteau08, paard10, masset10}. 
Migration is therefore expected to be inward and in the Type I regime.

Since the two-dimensional equations are obtained by averaging vertically, a softening length $\epsilon$ of order $h$ must be introduced in the gravitational potential of the planets. We take $\epsilon/h=0.6$ for both planets. Decreasing the softening length leads to an increase in migration speed and increases the relative strength of the corotation torque \citep{drag}. A value around $\epsilon/h=0.6$ gives good agreement with three-dimensional simulations \citep{muller12}. The planets do not have a physical radius in the simulations: they are treated as point masses (which are softened for the planet-disk interaction, but not for the planet-planet interaction). We can therefore not detect physical collisions, but we check afterwards the minimum distance between the planets to decide whether a collision has taken place.

%------------------------------
% Stochastic forces text
%------------------------------

\subsection{Stochastic forces}
\label{secNumStoch}

Even with state of the art computing clusters, high computational costs make it impossible to run fully turbulent simulations (for example due to the MRI) for $10^{4}$ orbits, which would be necessary to follow the migration of low-mass planets in discs with masses comparable to the MMSN. Therefore, we adopt a simplified description of the effects of turbulence on the evolution of the orbital parameters in terms of stochastic forces \citep{rein10}. Given an autocorrelation time $\tau_c$, a discrete first order Markov process can be used to generate correlated continuous noise which is then added to the force on the planets. The natural scale for such a force is the gravitational force per unit mass due to a small circular patch of radius $r_\Sigma$ at a distance  $\sqrt{2}r_\Sigma$ from the planet. Note that this scale, $F_0=\pi G\Sigma/2$, is independent of $r_\Sigma$ \citep{rein10}. A similar scale was used in \cite{oishi2007}. The correlation time is taken to be $\tau_c(r)=\Omega(r)^{-1}$, and we vary the amplitude $F$ of the correlated noise from $F/F_0=0.01$ to $F/F_0=0.1$. These values have been calibrated to mimic the forces measured in simulations of MRI \citep{oishi2007}. They measured the force per unit mass due to MRI turbulence on a fiducial planet in local simulations of the MRI for different values of the magnetic Reynolds number. However, it is important to note that these values are still uncertain, and measuring the strength of MRI turbulence is a very active field of research. Most recently, effects of non-ideal MHD are beginning to be studied \citep{simon13}.  Current MRI simulations either simulate the entire disc with low resolution, or a local patch with high resolution. The agreement between these fundamentally different sets is only marginal, and a very high resolution is needed to get comparable results \citep{sorathia12}. The results depend furthermore on many physical properties of the disc such as metallicity and ionization fraction. As a consequence one has to keep in mind that the value for $F/F_0$ might change by more than an order of magnitude. 

The Type I migration torque $\Gamma$ is proportional to $(q/h)^2 \Sigma r^4\Omega^2$ \citep{tanaka02,horse}, corresponding to a force per unit mass $F_\mathrm{mig}\sim (q/h^2) G\Sigma$. This means that even the smallest noise level we consider ($F/F_0=0.01$) gives rise to forces comparable to disc tides. 

%---------------------------------------------
% Hydrodynamical simulations text
%---------------------------------------------

\subsection{Results}
\label{secResHydro}

%----------------------------
% Laminar disc figures
%----------------------------

\begin{figure}
\centering
\resizebox{\hsize}{!}{\includegraphics[bb=0 0 253 347]{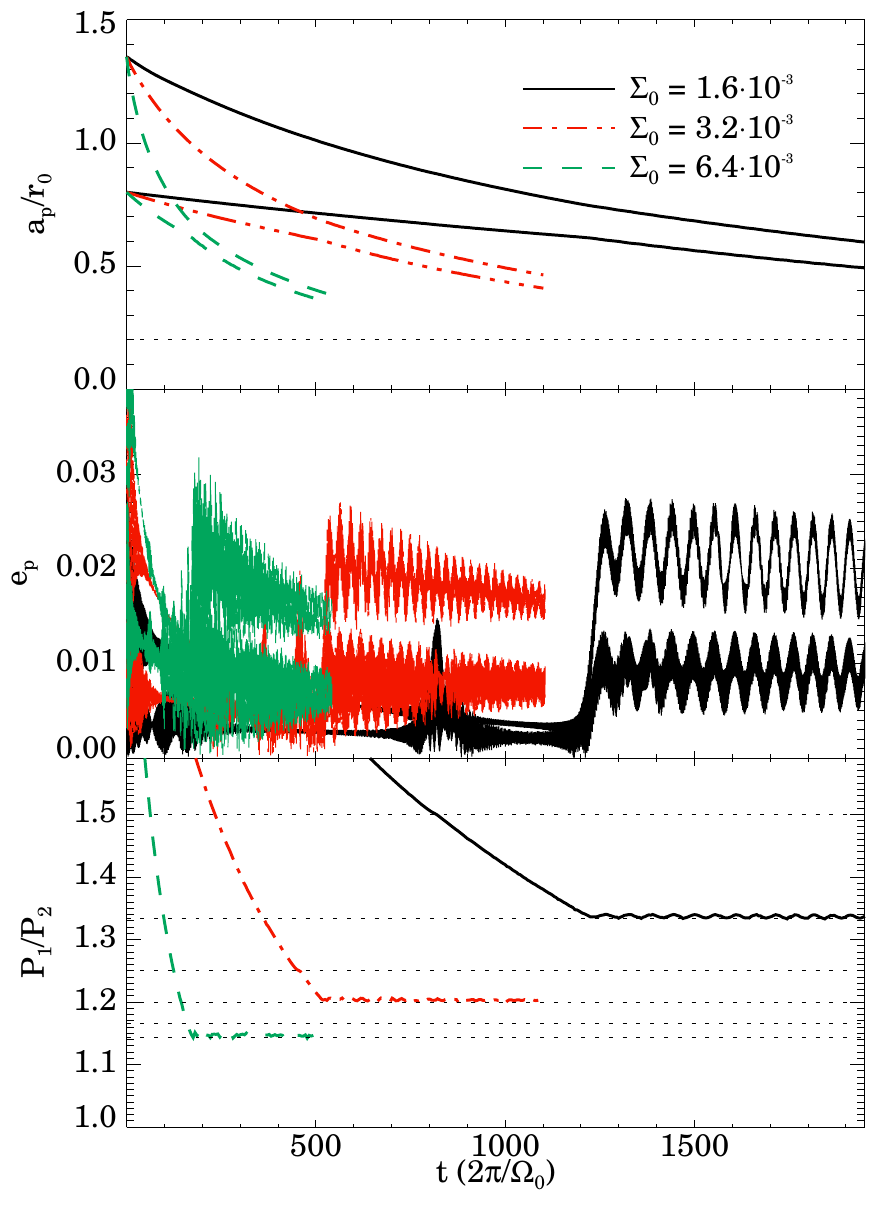}}
\caption{Evolution of a planetary system with masses similar to the Kepler-36 system in a hydrodynamic simulation without stochastic forces, for three different values of $\Sigma_0$ that refer to 8, 16, and 32 times the surface density of the MMSN. 
Note that for $r_0=1$ AU, time is in units of years. 
Top panel: semi-major axis; the horizontal dotted line shows the inner edge of the computational domain. 
Middle panel: eccentricity, where the largest value of $e$ always belongs to the inner, lower mass planet. 
Bottom panel: period ratio; the horizontal dotted lines show first order commensurabilities, from $3$:$2$ (top) to $8$:$7$ (bottom).}
\label{fig_laminar_aep}
\end{figure}

\begin{figure}
\centering
\resizebox{\hsize}{!}{\includegraphics[bb=0 0 251 345]{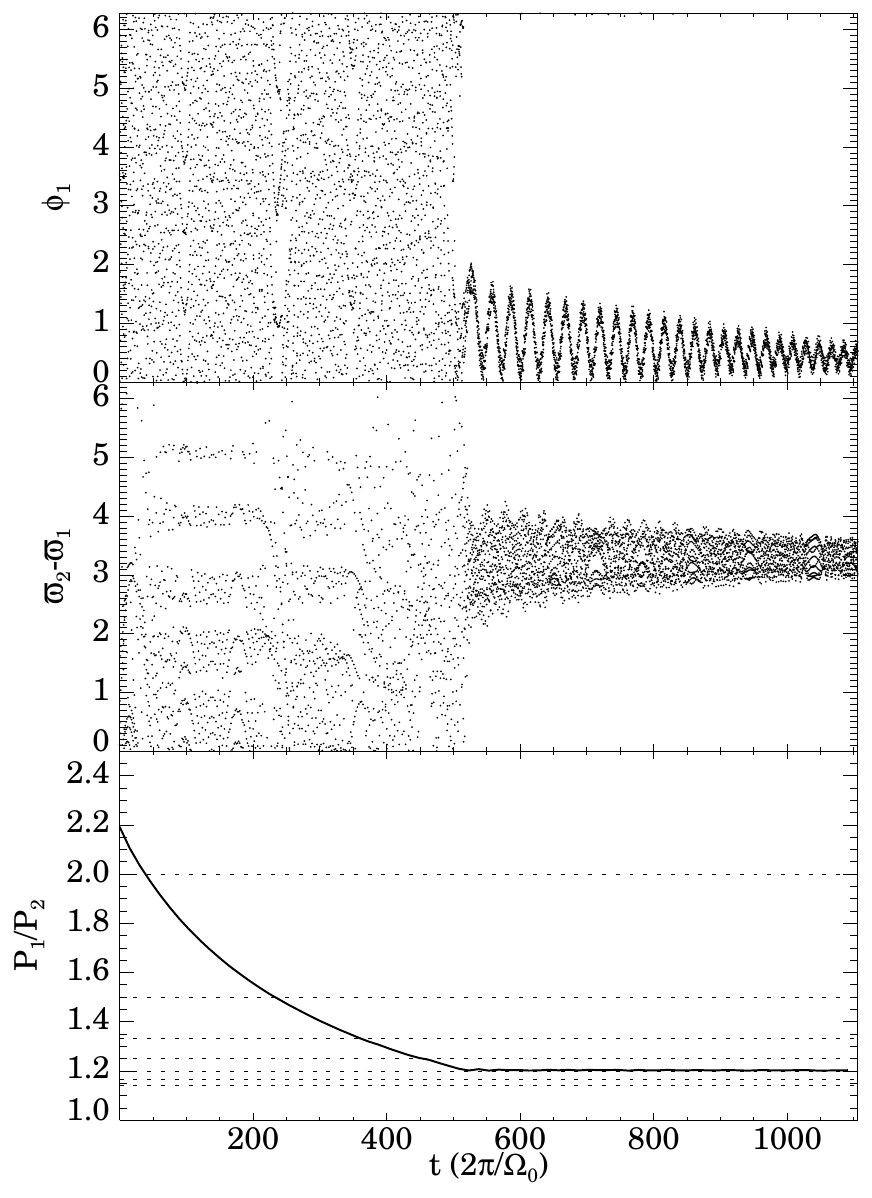}}
\caption{Evolution of the resonant angle $\phi_1$ for the $6$:$5$ resonance (top panel), difference between longitudes of periastron (middle panel) and period ratio (bottom panel) for a planetary system with masses similar to the Kepler-36 system in a hydrodynamic simulation without stochastic forces for $\Sigma_0=3.2\cdot 10^{-3}~M_*/r_0^2$. 
Note that for $r_0=1$ AU, time is in units of years.}
\label{fig_laminar_ang}
\end{figure}

%------------------------
% Laminar disc text
%------------------------

\subsubsection{Laminar discs}
\label{secResHydroLam}

We first consider the case without stochastic forces. 
This case was also considered in \cite{papaloizou05}, but without the specific system Kepler-36 in mind. 
We place the outer planet at $r/r_0=1.35$ and the inner planet at $r/r_0=0.8$, which puts them outside of the $2$:$1$ MMR, and varied the surface density $\Sigma_0$. 
We have found that for all surface densities considered ($\Sigma_0\geq2\cdot 10^{-4} M_* /r_0^2$, the MMSN surface density at 1 AU), the planets migrate through the $2$:$1$ resonance, in agreement with the results of \cite{papaloizou05} and our $N$-body simulations in Sect.~\ref{secNbody}. 
For $\Sigma_0 \leq 8\cdot 10^{-4} M_* /r_0^2$, the planets end up in the $3$:$2$ MMR. 
Obtaining a closely packed system requires a disc that is significantly more massive than the MMSN. 
The results for the highest disc masses, varying from $8$ to $32$ times MMSN, are displayed in Fig.~\ref{fig_laminar_aep}. 

For $\Sigma_0=1.6\cdot 10^{-3} M_* /r_0^2$, the planets end up in the $4$:$3$~MMR (see the bottom panel of Fig.~\ref{fig_laminar_aep}). 
Disc migration is strong enough to push them through the $3$:$2$~MMR around $\Omega_0 t /2\pi= 850$, an event that is recorded as a spike in the eccentricity evolution in the middle panel of Fig.~\ref{fig_laminar_aep}. 
After getting caught in the $4$:$3$~MMR, the eccentricities of the planets go up, but only to about $0.01$ for the outer, more massive planet, and $0.02$ for the inner planet. 
From the top panel of Fig.~\ref{fig_laminar_aep}, we see that the system continues to migrate inward at a rate $2\cdot 10^{-4} r_0\Omega_0/2\pi$. 
If we take $r_0=1$ AU, the associated migration time scale would be only $\sim5000$~yrs. 
This is alarmingly short, but we need even higher disc masses, and therefore faster migration rate, to come close to the $7$:$6$ MMR. 

When we further increase the disc mass by a factor of 2 ($\Sigma_0=3.2\cdot 10^{-3} M_* /r_0^2$), we find that the planets get locked into the $6$:$5$ resonance (bottom panel of Fig.~\ref{fig_laminar_aep}). 
Again, the orbital eccentricities increase, but remain smaller than $0.02$, in spite of the stronger forcing from the resonance because of the stronger eccentricity damping by the disc. 
Inspection of the resonant angles (see Fig.~\ref{fig_laminar_ang}) reveals that this configuration is stable. 
While $\phi_1$ oscillates around a value close to zero, $\phi_2$, and therefore $\phi_2-\phi_1=\varpi_2-\varpi_1$, oscillates around a value close to $\pi$, and the oscillations are decreasing in amplitude. 
We note that the same is true for the $4$:$3$ resonance discussed above. 

For $\Sigma_0=6.4\cdot 10^{-3} M_* /r_0^2$, the system attains a $8$:$7$~commensurability (bottom panel of Fig.~\ref{fig_laminar_aep}). 
Inspection of the resonant angles reveals that this configuration is stable as well. 
Therefore, it appears that in order to obtain a configuration close to the $7$:$6$~MMR, such as Kepler-36, we need a value of $\Sigma_0$ that is somewhere in between the highest two disc masses we considered. 
If we take $r_0=1$ AU, this means that we need a disc that is roughly 20 times more massive than the MMSN. 
While such a massive disc is just about gravitationally stable (Toomre $Q\sim 4$ for $h=0.05$), the associated migration time scales are alarmingly short ($\sim 2000$~yrs). 
Moreover, such high disc masses are only expected in the very early stages of disc evolution \citep[e.g.][]{lin90}, which means the formation time scales of the planets would have to be very short. 
Furthermore, questions are raised as to why none of the planets, while embedded in a very massive gas disc, was able to attract a significant amount of gas from the disc and thereby become a gas giant planet.   

For first-order resonances $p$+$1$:$p$, \cite{papaloizou05} observed that the dynamics had a stochastic character for high values of $p$, with some resonances becoming unstable on longer time scales. 
We have not observed such behaviour for our disc and planet masses before the planets reach the inner edge of the grid, which may be due to the different mass ratios considered here. In any case, we clearly need a very massive disc to come close to the $7$:$6$ resonance. However, the exact disc mass needed will certainly depend on the adopted surface density and temperature profile, since these are known to affect the Type I migration rates of planets \citep{paard10}. 
It should also be kept in mind that unless $\Sigma\propto r^{-2}$ migration is not scale-free. 
For surface density profiles that are less steep than this, migration time scales in terms of the dynamical time scale will \emph{increase} when a planet migrates inward. 
This suggests that closely-packed systems are more easily formed at large distances. 
Note, however, that this effect is rather small for a MMSN profile, for which $\Sigma_\mathrm{0,MMSN} \propto \sqrt{r_0/\mathrm{AU}}$. 
Based on our estimate of the necessary surface density $\Sigma_0$ for a $7$:$6$ resonance to be established, this particular resonance could have formed in a MMSN disc only at $400$~AU. 
For less steep density profiles the situation would be less extreme. 
Note that as long as migration remains convergent, a given resonance can always be maintained in principle, despite migration becoming relatively less strong. 

\cite{papaloizou05} also noted a dependence on initial conditions. 
This is most likely the result of incomplete eccentricity damping between resonances. 
If the planets reach a $p+1$:$p$ resonance with considerable leftover eccentricities obtained in the $p$:$p-1$ resonance, the chances of moving through the $p+1$:$p$ resonance increase \citep{kary93}. 
This effect becomes more important for higher values of $p$, for which the resonances are spaced closely together. 
It is interesting to note that, as far as this effect is concerned, in order to form a closely packed system it can be advantageous to start the planets farther apart. 
If we want the planets to move through the $6$:$5$ resonance, it is better to start the planets outside the $5$:$4$ resonance, so that they arrive at the $6$:$5$ with some eccentricity, rather than starting on circular orbits just outside the $6$:$5$ resonance. 
Regarding this issue, it should be noted that the initial planet orbits in {\sc fargo} are circular orbits if they could feel only the gravitational force from the central star. 
Adding the gravitational force due to the disc leads to some initial eccentricity, which depends on the disc mass and can be seen at early times in the middle panel of Fig.~\ref{fig_laminar_aep}. 
This eccentricity is subsequently damped by interaction with the disc, but starting the planets too close to a resonance could lead to spurious breaking of the resonance due to this initial eccentricity.

%----------------------------------
% Stochastic forces figures
%----------------------------------

\begin{figure}
\centering
\resizebox{\hsize}{!}{\includegraphics[bb=0 0 262 347]{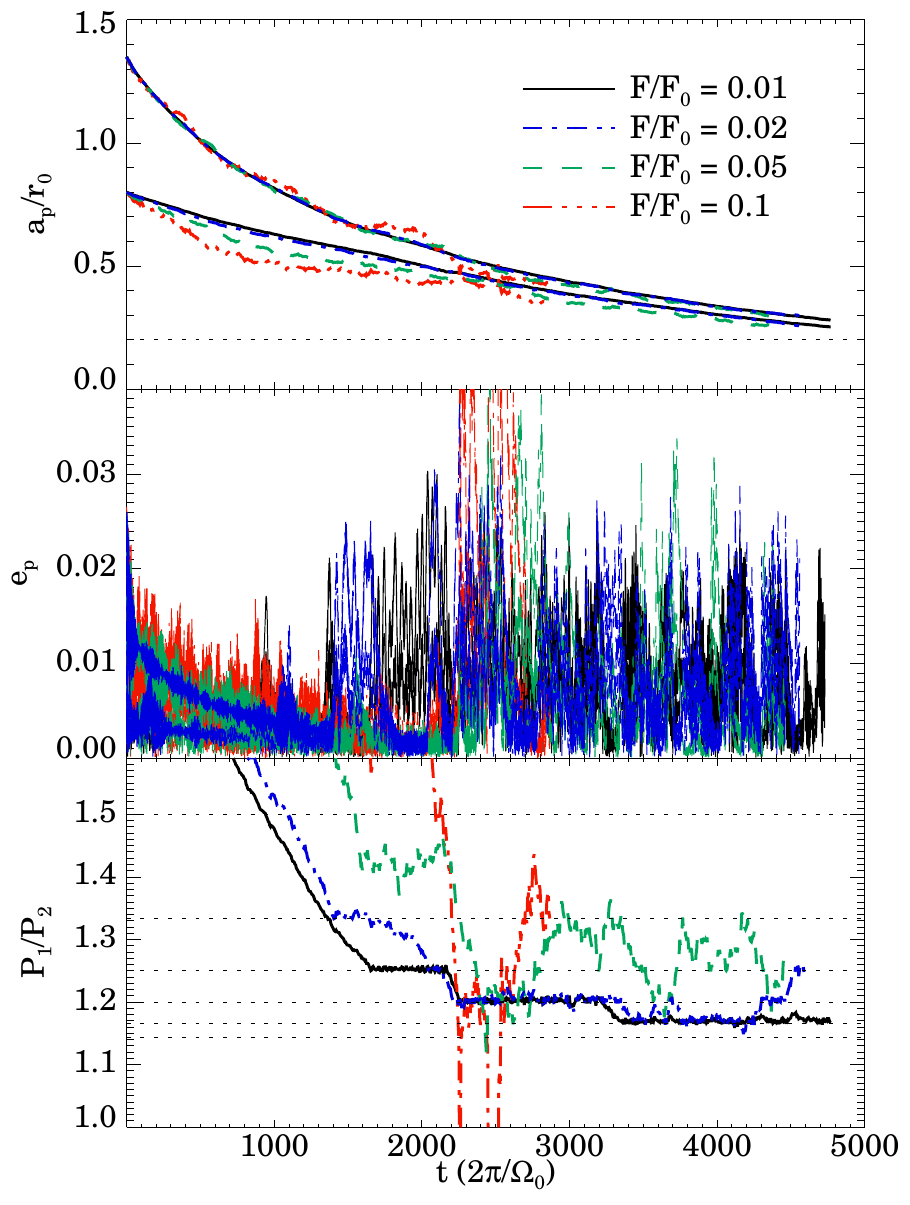}}
\caption{Evolution of a planetary system with masses similar to the Kepler-36 system in a hydrodynamic simulation with $\Sigma_0=1.6\cdot 10^{-3}$ with different levels of stochastic forcing. 
Note that for $r_0=1$ AU, time is in units of years. 
Top panel: semi-major axis; the horizontal dotted line shows the inner edge of the computational domain. 
Middle panel: eccentricity, where the largest value of $e$ always belongs to the inner, lower mass planet. 
Bottom panel: period ratio; the horizontal dotted lines show first order commensurabilities, from $3$:$2$ (top) to $8$:$7$ (bottom).}
\label{fig_stoch_16-3_aep}
\end{figure}

\begin{figure}
\centering
\resizebox{\hsize}{!}{\includegraphics[bb=0 0 251 345]{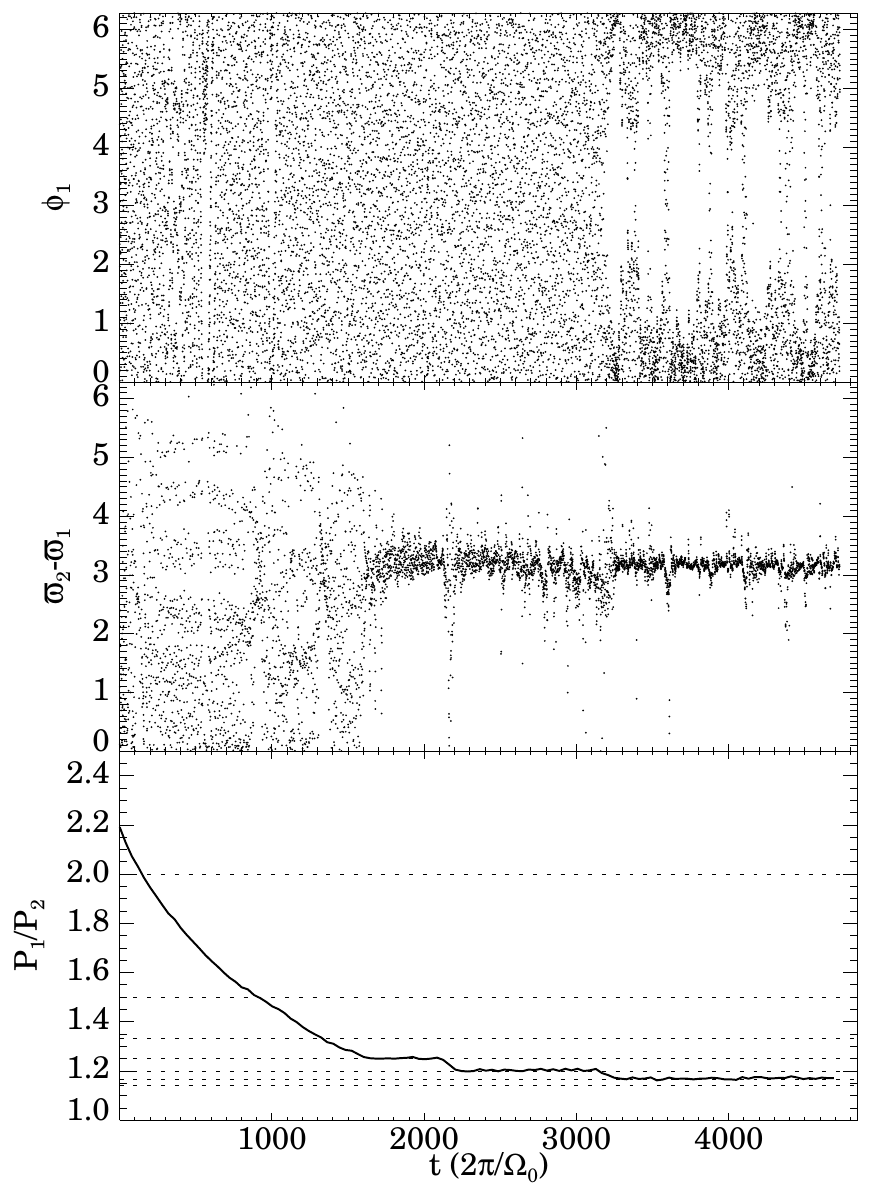}}
\caption{Evolution of the resonant angle $\phi_1$ (for the $7$:$6$ resonance; top panel), the difference between longitudes of periastron (middle panel) and period ratio (bottom panel) for a planetary system with masses similar to the Kepler-36 system in a hydrodynamic simulation with stochastic forces of amplitude $F/F_0=0.01$ for $\Sigma_0=1.6\cdot 10^{-3}~M_*/r_0^2$. 
Note that for $r_0=1$ AU, time is in units of years.}
\label{fig_stoch_16-3_ang}
\end{figure}

\begin{figure}
\centering
\resizebox{\hsize}{!}{\includegraphics[bb=0 0 251 345]{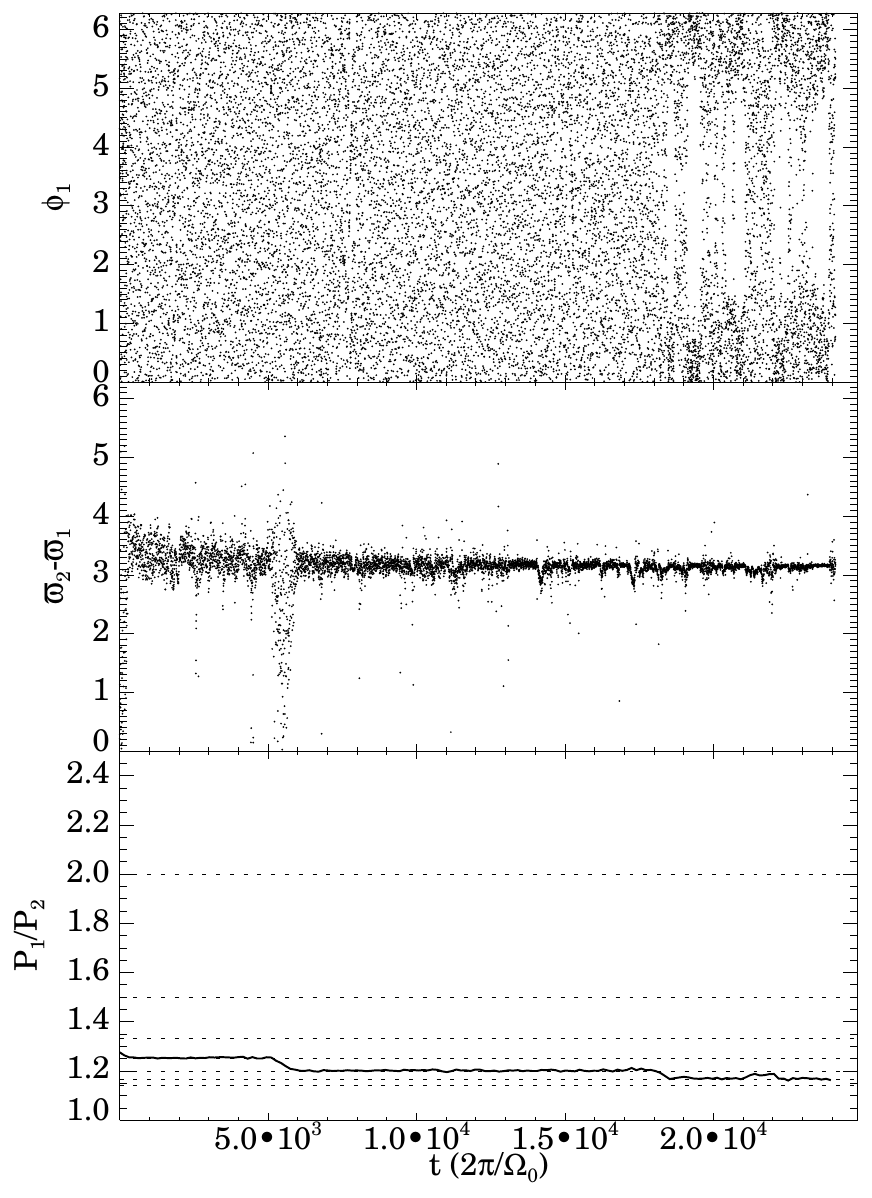}}
\caption{Evolution of the resonant angle $\phi_1$ (for the $7$:$6$ resonance; top panel), difference between longitudes of periastron (middle panel) and period ratio (bottom panel) for a planetary system with masses similar to the Kepler-36 system in a hydrodynamic simulation with stochastic forces of amplitude $F/F_0=0.02$ for $\Sigma_0=4.0\cdot 10^{-4}~M_*/r_0^2$. 
Note that for $r_0=1$ AU, time is in units of years.}
\label{fig_stoch_4-4_ang}
\end{figure}

%-----------------------------
% Stochastic forces text
%-----------------------------

\subsubsection{Effect of stochastic torques}
\label{secResHydroStoch}

We now let $F/F_0 \neq 0$, and consider first the case of $\Sigma_0=1.6\cdot 10^{-3}~M_*/r_0^2$, which was shown above to lead to a $4$:$3$ resonance in absence of stochastic forces. 
In Fig.~\ref{fig_stoch_16-3_aep} we show the results for four different amplitudes of stochastic forcing. 
For $F/F_0 \leq 0.02$, the semi-major axis evolution still appears relatively smooth (top panel of Fig.~\ref{fig_stoch_16-3_aep}), while for higher amplitudes stochastic forces visibly start to affect migration. 
Note that for $F/F_0=0.1$, the planets temporarily swap places around $\Omega_0t/2\pi=2500$.

Even though the migration rates of the individual planets appear unaffected by stochastic forces for $F/F_0\leq 0.02$, they have a profound effect on resonance locking. 
From the bottom panel of Fig.~\ref{fig_stoch_16-3_aep}, we see that even the lowest level of stochastic forcing moves the planets straight through the $4$:$3$ resonance. 
They briefly settle into $5$:$4$, but this resonance is also unstable in the presence of stochastic forcing, and after spending some time in $6$:$5$, the planets end up in the $7$:$6$ resonance before they come too close to the edge of the computational domain and we have to stop the simulation. 
Inspection of the resonant angles (see Fig.~\ref{fig_stoch_16-3_ang}) clearly shows the effects of the stochastic kicks experienced by the planets. 
While $\phi_1$ is close to zero on average, and $\phi_2$ close to $\pi$, there is no clear libration visible. 
This is to be expected, since stochastic forces continuously try to take the planets out of resonance. 
While the resonance may resist inward kicks, if the planets experience a kick that increases their period ratio, they are out of resonance, eccentricity damping sets in, and the resonant angles start to rotate. 
Convergent migration subsequently takes the planets back into resonance. 
Since the planets are never deep into a particular resonance in the presence of stochastic forcing, their eccentricities remain low on average, lower even than when $F/F_0=0$. 

As soon as the system spends some time in a resonance, the longitudes of periastron will be anti-aligned (see the middle panel of Fig.~\ref{fig_stoch_16-3_ang}). 
If the system gets kicked out of resonance, the resonant angles will start to rotate.
As soon as the next resonance is reached, $\varpi_2-\varpi_1$ will tend to $\pi$ again. 
When the resonances are closely spaced, as is the case for high values of $p$, this essentially means that the longitudes of periastron remain anti-aligned, which can be seen for $\Omega_0 t/(2\pi)\gtrsim 2500$ in the middle panel of Fig.~\ref{fig_stoch_16-3_ang}.

The presence of stochastic forces allows for closely spaced planetary systems in discs comparable to the MMSN. 
This is illustrated in Fig.~\ref{fig_stoch_4-4_ang}, where we show the resonant angles in a disc with $\Sigma_0=4.0\cdot 10^{-4}~M_*/r_0^2$ (twice the value of the MMSN if $r_0=1$ AU). 
Because of the long migration time scale in this low-mass disc ($\Omega_0\tau_a/(2\pi)\sim 10^5-10^6$), this simulation was done in several steps. 
As soon as the planets came too close to the inner boundary, we stopped the simulation, and put the planets back on their original position but at the period ratio they had before the restart. 
Care must be taken not to restart too close to a resonance, or the system may artificially move through the resonance because of the initial conditions (as was discussed at the end of Sect.~\ref{secResHydroLam}). 
Time in Fig.~\ref{fig_stoch_4-4_ang} is measured since the final restart, at which point were just outside the 5:4 resonance. 
After spending $\sim 5000$ yr (for $r_0=1$ AU) in the 5:4 resonance, the system moves into the 6:5, and subsequently the 7:6. 
After $\sim 21000$ yr, the system experiences an outward kick, but convergent migration is able to re-establish the 7:6 resonance after $\sim 1000$ yr. 

For $F/F_0>0.02$ different effects can be seen. 
Migration paths of individual planets become more erratic (see top panel of Fig.~\ref{fig_stoch_16-3_aep}), and consequently migration is no longer always convergent. 
Commensurabilities are no longer readily established, but closely spaced systems can still be formed. 
On average, migration is always convergent, pushing the system towards period ratios of order unity, while the strong resonances at high $p$ act as a barrier. 
As a result, for $F/F_0=0.05$, the system hovers between the 4:3 and the 8:7 resonance for $\Omega_0 t/(2\pi) > 2200$ in Fig.~\ref{fig_stoch_16-3_aep}. 
The strongest stochastic forcing we consider, $F/F_0=0.1$, allows the planets to switch places, which happens around $\Omega_0 t/(2\pi) \sim 2500$. 
This hovering between resonances is observed for all surface densities considered as long as $F/F_0>0.02$. 

While we can not detect physical collisions during the simulation, we can check afterwards whether a collision has taken place. We have measured the minimum distance between the planets for all simulations, and found that the run with $F/F_0=0.1$ the planets came as close as 12 $R_{p}r_0/\mathrm{AU}$, where $R_p=3.679$ $R_\oplus$ is the radius of planet c in the Kepler-36 system. Depending at which orbital distance this encounter took place (which determines the value of $r_0$ that should be used), it may have resulted in a physical collision, especially if the planets were larger in the past \citep{lopez13}. This minimum distance was reached when the planets switched places, which is an extreme event. For all other values of $F/F_0$, the minimum distance between the planets was always larger than 150 $R_{p}r_0/\mathrm{AU}$, making a physical collision far less likely.

It is important to point out that in the case of a close encounter, the result is a physical collision, not an ejection from the system.
This is because the escape speed from the planets' surface is less than $20~\text{km/s}$.
As mentioned about, if the planets were larger in the past, this value would be even smaller. 
On the other hand, the required speed for an ejection from the system is of the order of $130~\text{kms/s}$ (assuming the observed orbital parameters). 
Thus, the kick that the planets can get from a close encounter is not strong enough for either planet to leave the system.

%-----------------
% Results text
%-----------------

\section{$N$-body simulations}
\label{secNbody}

%---------------------------------
% $N$-body simulations text
%---------------------------------

\subsection{Numerical method}
\label{secNumNbody}
In addition to the hydrodynamical simulations described above, we also run $N$-body simulations to study the evolution of the Kepler-36 system.
In comparison to the hydrodynamical simulations discussed in Sect.~\ref{secNumHydro}, $N$-body simulations run much faster and we can explore a wide range of parameters.

We use the freely available {\sc rebound}\footnote{\url{https://github.com/hannorein/rebound}} code \citep{ReinLiu2012}.
The planets are setup in exactly the same way as in the hydrodynamical simulations, i.e. just outside the 2:1 resonance.
We add dissipative forces to the equations of motion which mimic the interactions of a planet with a proto-planetary disc \citep{LeePeale01}.
This allows us to choose two dimensional parameters for each planet, the migration timescale $\tau_a$ and the eccentricity damping time-scale $\tau_e$, which can be inferred from hydrodynamical simulations \citep{2007A&A...473..329C}. 
In all simulations presented here, we use an eccentricity damping timescale which is a factor of $K=100$ shorter than the semi-major axis damping timescale. 
We apply both migration and eccentricity damping to the outer planet, but only the eccentricity damping to the inner planet.
This setup ensures convergent migration. 
Tests have shown that varying $K$ and the precise nature of the dissipative forces do not significantly change the results.
However, simulations without any eccentricity damping on the inner planet are more likely to become unstable.
This is a consequence of angular momentum conservation which makes the eccentricity of the undamped inner planet grow while the outer planet pushes it inwards.
Stochastic forces are added in the same way as in the hydrodynamical simulations (see Sect.~\ref{secNumStoch}). 
As a normalisation, we use a disc with a surface density of $\Sigma_0=1.6\cdot 10^{-3}~M_*/r_0^2$ which allows us to use the same notation ($F/F_0$) as in the previous section.

%---------------------------------
% N-body simulations text
%---------------------------------

\subsection{Results}
\label{secResNbody}

\begin{figure}
\centering
\resizebox{\hsize}{!}{\includegraphics[bb=0 0 251 180]{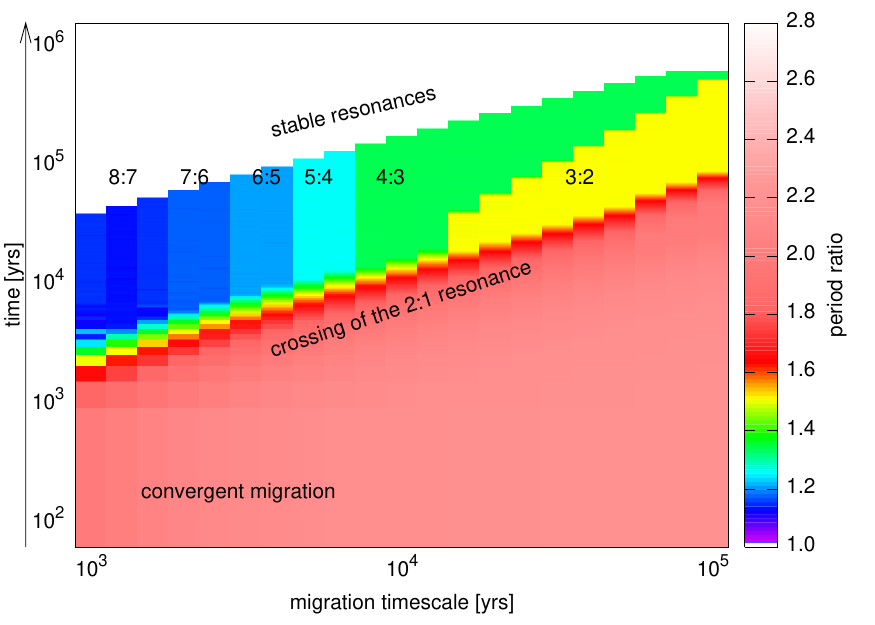}}
\caption{Results from $N$-body simulations with smooth migration forces. Colour represents period ratio. Time evolves upwards. Stable resonances form for migration rates longer than a few thousand years. Which resonance forms is determined by the migration rate.}
\label{fig_nbody_plot3}
\end{figure}

$N$-simulations allow us to confirm and understand the results of the hydrodynamic simulations with a much wider set of initial conditions.
We ran two sets of simulations, one with and one without stochastic forces.

The simulations without stochastic forces but with smooth migration are presented in Fig.~\ref{fig_nbody_plot3}. 
The plot shows a whole set of simulations.
The colour represent the period ratio $P_2/P_1$ of the planets.
Time is evolving upwards. 
Initially planets start outside the 2:1 resonance (as in the hydrodynamical simulations). 
Due to convergent migration, the period ratio shrinks. 
For all migration rates shown here ($\tau_a<10^5$~yrs) the planets pass through the 2:1 commensurability.
As one can see from this plot, in a smooth migration scenario, only extremely fast migration rates of the order of $1000$~yrs are able to form resonances of high $p$.

\begin{figure*}
\centering
\resizebox{\hsize}{!}{\includegraphics[bb=0 0 504 360]{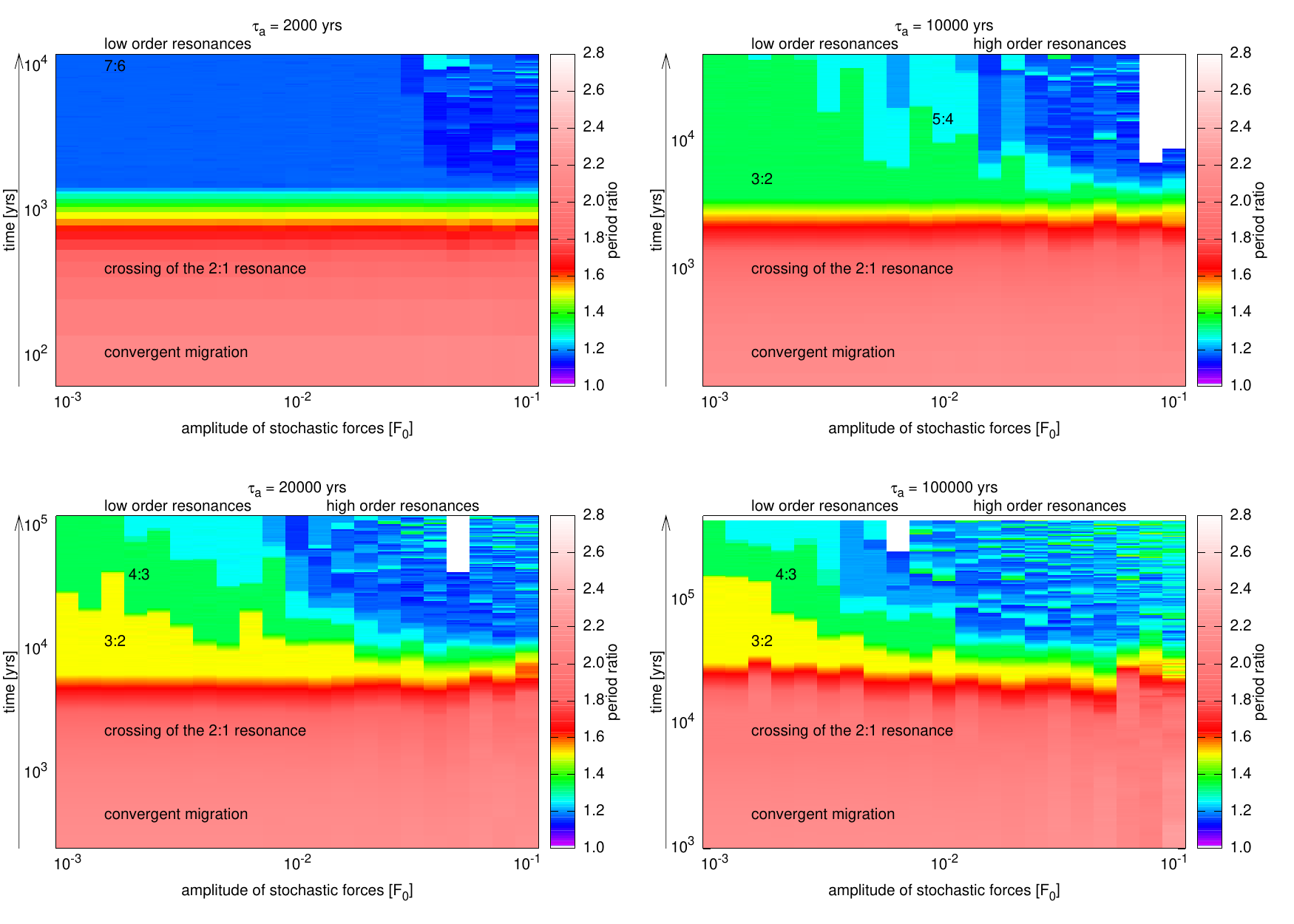}}
\caption{Results from $N$-body simulations with both smooth and stochastic migration forces. Colour represents period ratio. Time evolves upwards. Stable resonances form for migration rates longer than a few thousand years. The stochastic forces assume a disc with $\Sigma_0=1.6\cdot 10^{-3}~M_*/r_0^2$. For amplitudes $F/F_0$ larger than 0.01, resonances can break up and planets can capture into high-$p$ resonances.}
\label{fig_nbody_plot2}
\end{figure*}

The picture changes dramatically when stochastic forces are included. 
These results are presented in Fig.~\ref{fig_nbody_plot2}. 
As in Fig.~\ref{fig_nbody_plot3}, time evolves upwards and color represents period ratio.
The four panels correspond to four different migration rates, from top left to bottom right, $\tau_a = 2000~\text{yrs}$, $10000~\text{yrs}$, $20000~\text{yrs}$ and $100000~\text{yrs}$. 
For low amplitudes of the stochastic forces, $F/F_0<0.01$, the results are almost identical to those presented in Fig.~\ref{fig_nbody_plot3}. 
When stochastic forces are included low-$p$ resonances such as 3:2, 4:3 and 5:4 become unstable and the planets capture into resonances of higher $p$.
This can be seen by the dark blue colour on the top right of all four panels.
Note that even for moderate amplitudes of the stochastic forces, $F/F_0>0.01$, resonances break up and high-$p$ resonances can form. These resonances are not necessarily stable. 
As one can see in the plots, the precise resonance in which planets are in can change frequently (the color in the plots is not constant) when planets are in resonances of high $p$ and stochastic migration forces are present. However, the system itself remains stable and no planets are lost in most cases. 
Thus, when planets break out of a resonance, they re-capture into a (different) resonance as the convergent migration forces are still present.

This opens up a new route of formation for the Kepler-36 system. It is important to point out that it is a stochastic process. As one can see from the bottom right panel in Fig.~\ref{fig_nbody_plot2}, which of the high-$p$ resonances forms (7:6, 6:5, 5:4, 4:3) is highly unpredictable. However, it is important to point out that the systems tend to be in a resonance of high $p$ and stay stable for a long period. This in itself is a remarkable result as the parameter space contains large regions of instability and most orbits are chaotic. 
The interplay of stochastic and dissipative forces stabilises the system and prefers a specific set of orbital parameters. 

%---------------------------------
% Stability text
%---------------------------------

\section{Stability}
\label{secStability}

\begin{figure*}
\centering
\resizebox{\hsize}{!}{\includegraphics[bb=0 0 504 180]{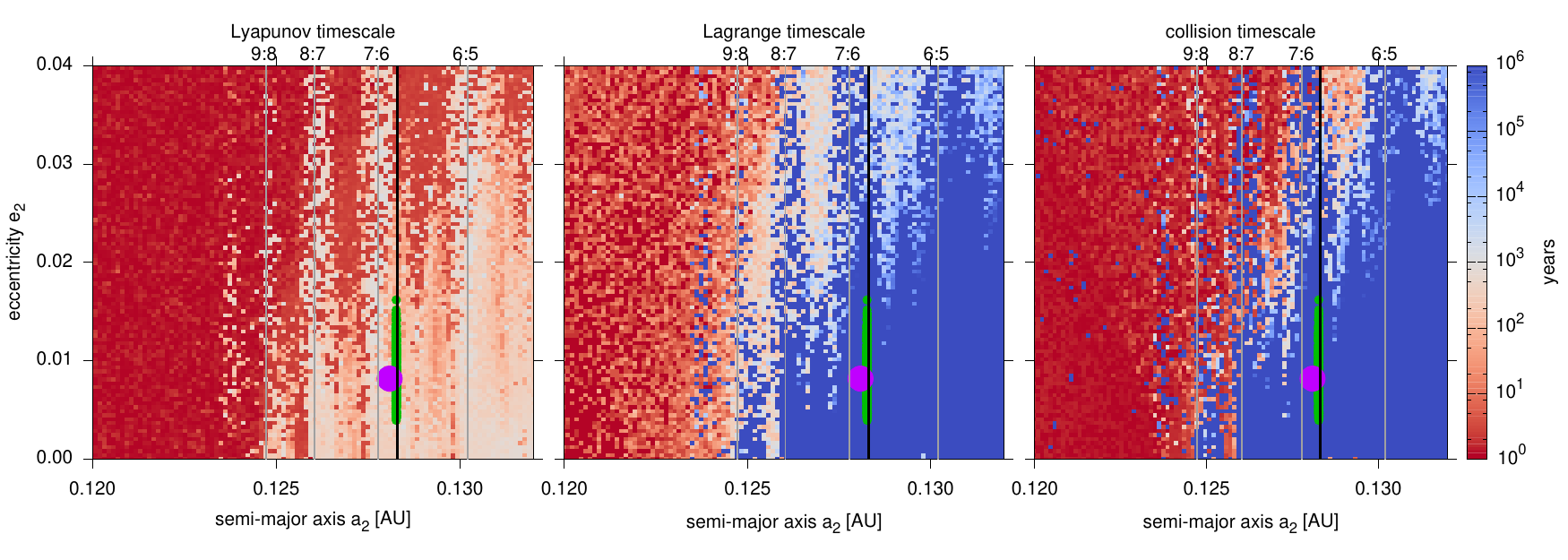}}
\caption{Different meassures of stability from long term $N$-body simulations. The maximum integration time is $10^6$~yrs. The inner planet is initially on a circular orbit with $a_1=0.1153$~AU. The $x$ and $y$ axes show the initial eccentricity and semi-major axis of the outer planet. The left panel shows the Lyapunov timescale of the system. The middle panel shows the Lagrange timescale. The right panel shows the time it takes for the planet to have a close encounter. The black line at $a_2=0.1283$ AU are solutions fitted to the observational data found by \citet{carter12}. Green dots represent the stable subset identified by \citet{deck12}. The purple dot is the final configuration of our hydrodynamical formation model.
}
\label{fig_stability1}
\end{figure*}

In this section we briefly discuss the stability of the observed Kepler-36 system and our orbital solutions.
A much more detailed analysis of the long term stability and the resonance overlap which can lead to chaotic orbits has been performed by \cite{deck12}.

\subsection{Measures of instability}\label{secMeasureOfInstability}
There are different measures that one can use to define wether a planetary system is stable or not which may lead to different answers.
In this paper, we define four different timescales that measure how long a system is stable.
\begin{enumerate}
\item The {\bf Lyapunov timescale} is a measure of how fast two nearby orbital solutions diverge. 
If the timescale is short, the system is called chaotic (i.e. sensitive to initial conditions). Often a chaotic system tends to be unstable. We will later see that this is not the case for the Kepler-36 system. See \cite{Wisdom1983} for more details on the numerical algorithm.
\item The {\bf collision timescale} measures the time until the two planets have a close encounter or a physical collision. 
In our simulations, we therefore measure the distance of the two planets at every timestep.

\item The {\bf ejection timescale} measures the time until at least one planet becomes gravitationally unbound from the host star and leaves the systems.
The Kepler-36 planets can only leave the system when they have a close encounter. 
This measure is therefore closely related to the Hill stability and the collision timescale of the system (see above). Note that for planetary systems of masses and densities comparable to the Kepler-36 system, instability usually results in a physical collision rather than ejection of one of the planets.

However, as \citet{deck12} point out, Hill stability is sufficient but not necessary for stability.
Our definition of the ejection timescale is purely based on direct measurements in $N$-body simulations.
\item Finally, the {\bf Lagrange timescale}, as for example used by \cite{deck12}, is the time until at least one of the planets' semi-major axis changes by more than 10\%. The planets might get ejected after they drastically changed their orbital parameters. However, as we will see, the more likely outcome for parameters similar to the Kepler-36 system is that the planet stay in a new configuration and remain stable from there onwards.
\end{enumerate}

\subsection{Overview of the parameter space}
We plot an overview of the phase space structure of the Kepler-36 system in Fig.~\ref{fig_stability1}. 
The $x$ and $y$ axes correspond to the outer planet's initial semi-major axis $a_2$ and eccentricity $e_2$, respectively.
The inner planet is initially on a circular orbit at a semi-major axis of $a_1=0.1153$~AU.
The system is coplanar and all angles are chosen from a uniform distribution.
We integrate the 10000~initial conditions shown in Fig.~\ref{fig_stability1} for $10^6$~years with the symplectic integrator of {\sc rebound} (see Sect.~\ref{secNumNbody}).
The color illustrates the different measures of stability defined in the previous Sect.~\ref{secMeasureOfInstability}:
The left plot shows the Lyapunov timescale, the middle plot shows the Lagrange timescale and the right plot shows the time until the planets have a close encounter.
We call an event a close encounter when the planets coming closer than $174 R_\oplus$
This distance is larger than the sum of the observed radii, $5.18~R_\oplus$ by a factor of 30. 
We chose this value because it allows us to use large timestep without missing any collisions or having to use an integrator with an adaptive timestep.
Other values have been tested and yield very similar results as long as the timesteps are not too large. The results were also compared to those obtained with an integrator with an adaptive time step (but run for only $10^5$ yrs), and both methods gave very similar results.

One can see that a large fraction of systems is long term stable in the sense that no planet gets ejected, collides or becomes Lagrange unstable (blue solutions in middle and right panels). 
However, the Lyapunov timescales are remarkably short, only a few hundred to a few thousand years (left panel).
This shows that these solutions, although Lagrange stable, are indeed chaotic \citep{deck12}.

There are a few minor caveats when comparing the orbital solutions in such a two dimensional plot as the parameter space is higher dimensional.
In particular, we are marginalising over the inner planet's eccentricity, the mass ratio, the mutual inclination, the longitudes and the periastrons.
We ran additional test which have shown that changes in the parameters which are fixed do not qualitatively change the structure of the phase space, but might very well shift the precise location of stable regions by a small amount. 
Nevertheless, Fig.~\ref{fig_stability1} gives us valuable information about the nature of the system.

\subsection{Orbital solution from observations}
The best fits to the transit light curves of Kepler-36 are shown in Fig.~\ref{fig_stability1} as a black line.
These solutions are the outcome of a MCMC regression by \cite{carter12} which does not take into account long term stability.
We scaled the semi-major axis such that the inner planet's semi-major axis is always at $a_1=0.1153$~AU. 
This is effectively fixing the stellar mass and allows us to over-plot these solutions in our parameter space survey.
Note that by doing so the semi-major axis of the outer planet is almost constant ($a_2\sim0.01283$~AU) and the set of solutions appears therefore as a vertical line. 
This is a result of Kepler measuring the period (and therefore the period ratio) of transits to high accuracy.

We also over-plot the long term stable solutions found by \cite{deck12} as green dots. 
These are a subset of the \cite{carter12} solutions. 
The stable solutions are outside the nominal $7$:$6$ commensurability and very close to the boundary where long term stable solutions can exist.

%-------------------------------
% FARGO solution table
%-------------------------------

\begin{table}
\centering
  \begin{tabular}{lrr}
  \hline
    & Planet b & Planet c \\ \hline
Semi-Major axis (AU) & $0.11530$ & $0.12809$ \\
Eccentricity & $0.014667$ & $0.008180$ \\
Longitude of periastron (rad) & $1.8586$ & $-1.2668$\\
True anomaly (rad) & $-1.2091$& $1.2825$\\
\hline
  \end{tabular}
  \caption{Orbital parameters derived from a hydrodynamical simulation with $\Sigma_0=4.0\cdot 10^{-4}~M_*/r_0^2$ and $F/F_0=0.02$ after the surface density has been decreased exponentially to zero.}
  \label{tabFARGO}
\end{table}

\subsection{Orbital solution from formation model}
We have taken the hydrodynamic simulation defined by $\Sigma_0=4.0\cdot 10^{-4}~M_*/r_0^2$ and $F/F_0=0.02$ shortly after the $7$:$6$ resonance is reached, and decreased the surface density exponentially over a time scale of $\Omega_0 t = 200\pi$ (100 years for $r_0=1$~AU). Thus, convergent migration as well as stochastic forcing is turned off slowly. The system then settles into an orbital configuration close to the $7$:$6$ resonance. 
We then set $r_0=0.336467$~AU so that the semi-major axis of the inner planet is matching the observed value of $0.1153$~AU. 
The resulting orbital parameters are summarised in Table~\ref{tabFARGO}. Since the effects of the gas disc are turned off, this solution can be continued with an $N$-body algorithm. Note that the semi-major axis ratio is consistent with the confidence interval given in \cite{carter12} (their table S7). 
  
The purple dot in Fig.~\ref{fig_stability1} corresponds to the {\sc fargo} solution presented in Table~\ref{tabFARGO}. This solution is stable for at least $10^6$ years. Remarkably, the Lyapunov timescale of this solution is very long ($> 10^5$ yrs). This suggests that the solution is not chaotic. \cite{deck12} report that only a small fraction ($<1\%$) of their solutions exhibit such long Lyapunov timescales. We can confirm this. All simulations in our parameter space survey have a short Lyapunov timescale (see left panel in Fig.~\ref{fig_stability1}). It is therefore even more remarkable that the stochastic migration scenario picks out a solution which is not only stable, but also not chaotic.

We further tested the robustness of this result by a series of $N$-body integrations randomly perturbing the orbital parameters given in Table \ref{tabFARGO} by $1\%$.
The results indicate that roughly $50\%$ of those solutions are also not chaotic. 
We are therefore confident that our solution has a finite measure in parameter space and can act as an attractor in formation scenarios such as the one presented here. 
Note however, that although the best fits from the observations and our solution coming from a formation scenario are very similar, they are not identical. While the semi-major axis ratio is consistent with the observations, inspection of the eccentricity parameters given in table S7 from \cite{carter12} reveals that the eccentricities of the hydrodynamic solution are slightly too low. We tested that, keeping the other orbital parameters fixed, there are still stable, non-chaotic solutions at higher eccentricities that are consistent with \cite{carter12}. It is possible that for a different value of $\Sigma_0$ or $F/F_0$, higher eccentricities can be reached. We have made no attempt to fine-tune the hydrodynamic outcome to match the observations exactly.  

%----------------------
% Conclusions text
%----------------------

\section{Conclusions}
\label{secCon}
In this paper, we have studied a possible formation scenario for closely spaced low-mass planets, like Kepler-36, through convergent migration in a turbulent disc. 

The most important result of our study is that it is indeed possible to form such a closely-packed system by convergent migration in the presence of stochastic forces. 
This is the first successful formation scenario for such a highly compact system of low mass planets. 
While a scenario involving smooth migration requires a very large disc mass, adding the effect of turbulent density fluctuations through stochastic forces onto the planets naturally lead to the formation of resonances of very high values of $p$. 
Hydrodynamic and modified $N$-body results give very similar results. 
We find that even though the planets get very close together and individual resonances may become unstable, the system itself remains surprisingly stable. 
When slowly removing the gaseous disc, the hydrodynamic simulations tend to pick out stable, non-chaotic solutions. 

In order to make the problem tractable computationally, several simplifications had to be made. First of all, we adopted a simple prescription for the effects of turbulence through stochastic forces acting on the planets. In a real disc, the turbulence is likely created by the MRI. Unfortunately, global MRI simulations spanning $10^5$ orbits or more, necessary to study the formation of the Kepler-36 system at low disc masses, are not feasible right now. We have tried to choose the parameters of our stochastic forcing, amplitude and correlation time, to be as close to those of the MRI as possible (see Sect.~\ref{secNumStoch}), but it is clearly necessary to work towards more realistic turbulent simulations. We leave this to future work. 

In a realistic disc, it is likely that the strength of turbulence as well as the migration rate vary strongly with distance to the star. For example, the magnitude and even the direction of Type I migration strongly depends on background temperature and density profiles of the disc \citep{paard10}, while the level of turbulence depends on the ionisation fraction \citep{gammie96}. Depending on where the planets first 'meet' and lock into resonance, their subsequent journey in the disc may lead them through various migration and turbulence regimes. While in some cases, these regimes may be such that the planets' orbits do not converge or even diverge, the final outcome is largely determined by the last phase of migration and turbulence experienced by the system.

We have worked in the 2D approximation, using only vertically integrated quantities like the surface density.
Using an appropriate value for the softening parameter in the planets' gravitational potential, it is possible to a large extent to reproduce the results of 3D hydrodynamical calculations as far as migration is concerned \citep{muller12}. 
Moreover, fully 3D hydrodynamical simulations for more than $10^5$ orbits are beyond current computational resources. 
This means that we can only consider a restricted region of parameter space, namely coplanar orbits. 
While the mutual inclination in Kepler-36 system can be constrained to be smaller than $2.5^{\circ}$ \citep{carter12}, making the system at least almost coplanar, a small relative inclination can affect the stability of the system \citep{deck12}. 
Such effects are probably best studied using the $N$-body approach. 

A formation scenario involving migration needs to somehow address the question of how and at what point migration is stopped. 
One possibility for low-mass planets such as in the Kepler-36 system is the existence of a `planet trap', where temperature and/or density gradients are such that the Type I torque changes sign. 
It remains to be seen in what way trapping the inner planet in such a way changes the current picture.   

We have kept the mass of the planets fixed. 
If the different densities of the planets of Kepler-36 came about through a different formation location, it is necessary for the outer planet to obtain its gaseous envelope before the planets come close together. 
The early evolution of the system will strongly depend on the mass evolution of the outer planet.
For example, if the solid core of the Neptune-type planet is less massive than the inner Super-Earth, convergent migration will naturally only commence once the outer planet has obtained its envelope. 

We also ignore the effects due to strong X-ray and EUV irradiation on the planets. 
The work of \cite{OwenWu2013} shows that the inner planet in the Kepler-36 system might have undergone a phase of evaporation. If the mass-loss was significant, it might have changed the orbital configuration of the planets after the protoplanetary disc had long disappeared. The mass-loss is stronger for planets closer to the star and might therefore provide an alternative explanation for the different observed densities in the Kepler-36 system. Evaporation was also considered in \cite{lopez13}, who argue that a different core mass for the planets can lead to the density contrast observed today. It will be interesting to see if such a scenario can be combined with converging migration to provide a full history of Kepler-36.

We have not fully explored the possibility of physical collisions in the system during its evolution. 
While in most of our simulations, the planets remain at safe distances from each other, the largest amplitude of stochastic forcing, for which the planets switch places, leads to a close encounter that may have resulted in a physical collision. 
Such an event is more likely if the planets were larger in the past \citep{lopez13}. 
Since the two-dimensional nature of our hydrodynamical simulations requires the use of a smoothing length in the planet potential, studying the effects of these close encounters in a gas disc in more detail will require three-dimensional simulations. 
It is also likely that the frequency of these events will be reduced in a three-dimensional setup.

It is worth pointing out that Kepler-36 is an intrinsically rare system. Systems of such short periods and small period ratio have a high probability of being seen in transit. Therefore, even if a specific setup is needed in order to form this system, or even if the system spends relatively little time in the $7$:$6$ resonance (e.g. for $F/F_0=0.02$ in figure \ref{fig_stoch_16-3_aep}), this may still be an acceptable formation pathway.

There are many opportunities for future studies involving Kepler-36.
More advanced simulations simulations should allow for accretion onto the planet, model the irradiation from the star and provide a self-consistent stopping mechanism for the planets.
The orbital solution picked by our formation scenario is very close to, but not exactly, the observed one, there is the hope to exactly match the parameters. This will allow us to make testable predictions for poorly constrained orbital parameters, similar to what \cite{ReinPapaloizouKley2010} did for HD45364.

%----------------------------
% Acknowledgements 
 %---------------------------
 
\section*{Acknowledgements}
We would like to thank the referee Darin Ragozzine for comments which greatly improved this manuscript.
We would also like to thank Katherine Deck, Josh Carter, Yanqin Wu, James Owen and Cl\'ement Baruteau for stimulating discussions throughout this project.
SJP is supported by a Royal Society University Research Fellowship.
Hanno Rein was supported by the Institute for Advanced Study and the NFS grant AST-0807444.
Some of the simulations using {\sc fargo} were performed using the Darwin Supercomputer of the University of Cambridge High Performance Computing Service (http://www.hpc.cam.ac.uk), provided by Dell Inc. using Strategic Research Infrastructure Funding from the Higher Education Funding Council for England.

\bibliography{resonance}

\label{lastpage}

\end{document}